\documentclass{article}

\usepackage{arxiv}
\usepackage{amsmath}
\usepackage{float}
\usepackage[utf8]{inputenc} 
\usepackage[T1]{fontenc}    
\usepackage{hyperref}       
\usepackage{url}            
\usepackage{booktabs}       
\usepackage{amsfonts}       
\usepackage{nicefrac}       
\usepackage{microtype}      
\usepackage{cleveref}       
\usepackage{lipsum}         
\usepackage{graphicx}
\usepackage{doi}
\usepackage{amssymb}

\usepackage{algorithm,algorithmic}

\usepackage{natbib}

\usepackage{tikz}
\newcommand*\circled[1]{\tikz[baseline=(char.base)]{
            \node[shape=circle,draw,inner sep=2pt] (char) {#1};}}

\begin{document}
\title{Insights from the Application of Nonlinear Model Predictive Control to a Cart-Pendulum}


\author{Mark P. Balenzuela\\ \vspace{1mm}\texttt{Mark.P.Balenzuela@gmail.com}}
	

\newcommand\Authfont{\bfseries}
\newcommand\diag{\text{diag}}

\renewcommand{\shorttitle}{\textit{arXiv} Template}

\newcommand{\bU}{\mathbf{u}}
\newcommand{\bA}{\mathbf{A}}

\hypersetup{
pdftitle={A template for the arxiv style},
pdfsubject={q-bio.NC, q-bio.QM},
pdfauthor={David S.~Hippocampus, Elias D.~Striatum},
pdfkeywords={First keyword, Second keyword, More},
}

\maketitle

\begin{abstract}
Inspired greatly by \cite{mills2009nonlinear} and the solution within, this paper aims to more clearly explain the mathematics and implementation details of such a powerful control algorithm.
While the aforementioned paper is well written and of sound mathematics, it is extreamly dense and requires some time and patience to decipher, especially as it draws on many other sources to complete the algorithm.
This dense property is a clear result of the paper being restricted to the brief form and important details being ommited as a result. We provide the much needed elaboration here for the benifit of the reader.
\end{abstract}


\section*{Work in Progress!}
This is a work in progress and is not complete, and there may be typos, errors or inaccuracies in this document. I am writing this alongside a full-time job, so progress is going to be slow. Bare with me.

\section{Introduction}
Non-linear model predictive control (NMPC) is by nature capable of controlling many non-linear systems, where other control algorithms, especially linear schemes, would struggle greatly.
The downside of such a powerful scheme is a more complex algorithm which requires more time, patience and care to implement.
Sensitivity to both model parameters and NMPC tuning parameters such as horizon time may cause convergence to the setpoint to not occur, and appear to make NMPC less robust when compared to traditional methods.
Further downsides stem from the real-time implementation itself, as too slow of solve will compute the required action after it was due to be allocated to the actuators, and fail to control the plant.
For these reasons, it should not be a go-to algorithm, but one reserved for difficult control problems where extra effort is warrented.

Such examples commonly include constraints on input effort or valid operating region of the plants states.
Other indications include operation of the plant within areas where there is of non-monotonic behaviour and linearisation will provide an extreamly poor approximation of the function, such as a pendulum in the downward position.
It is precisly this use case, in which both the parent paper and this paper targets.

\subsection{The Jist}
Given the starting state of the system $x_0$, and a possible input sequence $u_{1:N}$, the state trajectory of the system can be predicted over a window in the future $x_{1:N}$, where the extent of this window is reffered to a Horrizon, $N$.
The input sequence $u_{1:N}$ and predicted state trajectory $x_{1:N}$ can be scored according to a cost function $V(u_{1:N}, x_{1:N})$, where such a function will typically reward minimal use of input effort and ideal trajectory of the states.
Searching of the valid regions of $u_{1:N}$ can then be completed to find the optimal input sequence $u_{1:N}^*$ that corresponds to the optimum outcome of state trajectory over the horzon, which by definition will minimise the cost function.

Upon finding the optimal sequence $u_{1:N}^*$, this sequence can begin to be issued to the system with the correct timing, while the state estimate $x_0$ is updated from new measurements $y_1$, and the search for the optimal sequence over a shifted horizon takes place. Note that to maintain a brief compact notation, we outline just one iteration of NMPC algorithm, and the shifted horizon to begin at the new origin $k=1$.

It would be extreamly disadvantagious to start the search for an optimal input sequence in the shifted horizon without using the solution from the previous horizon.
A good idea is to start this new search with the shifted sequence $u^+_{1:N}$, where the issued part of the optimal sequence is removed, and the optimal input at the horizon is duplicated the neccisary number of times, i.e., if the new optimal could be completed before the next input was due to be allocated, then the hotstart sequence would be $u^+_{1:N} = [ u^*_{2:N},  u^*_{N} ].$

By repeating this search for an opimal input sequence before allocating said sequence to the system, the system theoretically is controlled in an optimim way provided that the models behaviour does not deviate too greatly from the plant.
Furthermore, some errors in the model can be corrected by using measurements to update the state estimate at the start of the horizon, making this an example of closed-loop control.

\section{Quadratic Programming}

Quadratic Programming (QP) is simply a term used for solving a quadratic equation.
In the unconstrained case, this is most trivial to solve naively, e.g. for the scalar case
$$y=ap^2+bp+c, \quad p^*=\frac{-b}{2a},$$
and similarly for the multivariate case
\begin{align}
y = \frac{1}{2}p^THp+\vec{g}^Tp+c, \quad p^* = -H^{-1}\vec{g}.
\label{eq:quadraticbase}
\end{align}

\subsection{Optimising using Sequential Quadratic Programming}
The general idea here is: if we are at a point on a surface, and we have the gradient and hessian for this point, we could approximate the local region as a quadratic surface with the origin at our current positon.
Interestingly, an evaluation of the function at the point is not required to constuct such a surface as the solution is not dependant on the $c$ term.
We could then use the solution to the quadratic problem to move on the original surface, either by:
\begin{itemize}
\item Taking a step on the original surface with an equal length and direction to the quadratic solution, or
\item Conducting a line search on the original surface in the direction given by the solution to the quadratic surface.
\end{itemize}
The former option, while attractive due to its simplicity, may overshoot the lower point should the local quadratic approximation be poor. Additionally, a line search with constraints becomes more complex again.

\subsection{Constraints}
Equality constraints effectively reduce the space to be searched, and if leveraged correctly, can be efficient to solve, but with the additon of inequality constraints, comes greater complexity as constraints may be active or inactive.
While the original paper does not specifically any detail of the specific method to solve the SQP problem with constraints, an interrior point method called Mehrotra's predictor–corrector method is well suited.

\subsection{A QP Solver, Mehrotra's Predictor–Corrector Algorithm}

The interested reader is encouraged to do a deep dive into numerical optimisation by reading the following text \cite{nocedal1999numerical}, specifically chapter 16.
Mehrotra's predictor–corrector algorithm originally was for a linear-program problem, but has since been expanded to solve QP.
For those wishing to skip this, a compact implementation of the Mehrotra's predictor–corrector algorithm for a QP is as follows, which is is an edited version of Algorithm 16.4 in this papers notation for simplicity.

For the problem $\frac{1}{2} p^THp +\vec{g}^Tp$
with the constraints $Ap \geq b$

\begin{align}
\Psi(\lambda,y) = \begin{bmatrix}H & \mathbf{0}_{\bar{n},m} & -A^T\\A&-\mathbf{I}_{m,m} & \mathbf{0}_{m,m}\\\mathbf{0}_{m, \bar{n}} & \diag (\lambda) & \diag (y)\end{bmatrix},
\end{align}

\begin{align}
\gamma(p,\lambda, y, \Delta \lambda^\text{aff} , \Delta y^\text{aff} ,\sigma, \mu) = \begin{bmatrix} A^T \lambda -  \vec{g} -Hp  \\ y - Ap +b\\-\lambda \odot y- \Delta\lambda^\text{aff} \odot \Delta y^\text{aff}+\sigma\mu \vec{\mathbf{1}}_m \end{bmatrix},
\end{align}

where $\bar{n}=Nn_u$, $m$ is the number of inequality constraints, and $\vec{\mathbf{1}}_m$ is a vector of length $m$ containing $1$ in all elements.

where $\odot$ is the Hadamard product\footnote{also called a Schur product.}, denoting element wise multiplication, and $\diag(\cdot)$ is the diagonal function, which places elements of a vector at index $(i)$ on the diagonals of a square zeros matrix at index $(i,i)$.

A key component of the algorithm is the calculation of the step-length scaling parameter. This involves solving the equation

\begin{align} 
\alpha^*(v,r) = \max\{\alpha \in (0,1]: \alpha v \geq r \} \label{eq:alphastar}
\end{align} 

For elements of $v$ where $v_i >0$ this equation is fairly trivially maximised with $\alpha^*=1$, and this is most likely to satify the inequality condition.
But for negative elements where $v_i<0$, an upper bound is placed on $\alpha^*$, as
\begin{align} 
\alpha v_i \geq r_i \\
\alpha \leq \frac{r_i}{v_i}
\end{align}

The minimum of these upper bounds is the solution to the maximisation problem \eqref{eq:alphastar} provided that the value is not greater than 1, in which case $\alpha^*=1$.

So if $v$ and $r$ are vectors of length $n$, then
\begin{align} 
\label{eq:solveAlphapls}
\alpha^*(v,r) = \min\{1, w(v_1,r_1), w(v_2,r_2), ..., w(v_n,r_n)\}, 
\quad
w(v_i,r_i) = \begin{cases} 
      1 & v_i \geq 0 \\
     \frac{r_i}{v_i} & v_i < 0
   \end{cases}
\end{align}

Mehrotra's Predictor–Corrector Algrithm for a quadratic problem is now detailed in Algorithm~\ref{alg:MetPredCorr}.
Note that it is explicity mentioned that $\tau$ could be a function approaching a value of $1$ as the solution converges, and that this can speed up convergence.

\begin{algorithm}[!htbp]
\begin{algorithmic}
\caption{Mehrotra's Predictor–Corrector}
\label{alg:MetPredCorr}
\REQUIRE Parameterisation of the QP problem by supplying $H$, $\vec{g}$, $A$, and $b$
\REQUIRE $n$, the number of iterations to use to solve the QP problem.
\REQUIRE A choice of tuning parameter $\tau = (0,1)$.
\REQUIRE Initialisation values for $y$, and $\lambda$. 
\STATE $p \gets \vec{\mathbf{0}}_{\bar{n}}$
\FOR{$i=1,..., n$}
	\STATE $\Delta \lambda^\text{aff} \gets \vec{\mathbf{0}}_m$
	\STATE $\Delta y^\text{aff} \gets \vec{\mathbf{0}}_m$
	\STATE $\sigma \gets 0$
	\STATE $\mu \gets 0$
	\STATE Backsolve for $\Psi^{-1}(\lambda,y) \gamma(p,\lambda, y, \Delta \lambda^\text{aff} , \Delta y^\text{aff} ,\sigma, \mu) $, and separate the resulting vector into $[\Delta p^\text{aff},\Delta y^\text{aff},\Delta \lambda^\text{aff}]$
	\STATE Using \eqref{eq:solveAlphapls}, $\alpha_p^\text{aff} \gets \alpha^*(\Delta y^\text{aff}, -y)$
	\STATE Using \eqref{eq:solveAlphapls}, $\alpha_d^\text{aff} \gets \alpha^*(\Delta \lambda^\text{aff}, - \lambda)$
	\STATE  $\alpha^\text{aff} \gets \min(\alpha_p^\text{aff}, \alpha_d^\text{aff})$
	\STATE $\mu^\text{aff} \gets (y+\alpha^\text{aff}\Delta y^\text{aff})^T(\lambda+\alpha^\text{aff} \Delta \lambda^\text{aff})m^{-1}$
	\STATE  $\mu \gets { y^T\lambda}{m^{-1}}$
	\STATE $\sigma \gets (\mu^\text{aff}\mu^{-1})^3$
	\STATE Backsolve for $\Psi^{-1}(\lambda,y) \gamma(p,\lambda, y, \Delta \lambda^\text{aff} , \Delta y^\text{aff} ,\sigma, \mu) $, and separate the resulting vector into  $[\Delta p,\Delta y,\Delta \lambda]$
	\STATE Using \eqref{eq:solveAlphapls}, $\alpha_p \gets \alpha^*(\Delta y, -\tau y)$
	\STATE Using \eqref{eq:solveAlphapls}, $\alpha_d \gets \alpha^*(\Delta \lambda, -\tau \lambda)$
	\STATE $\alpha \gets \min(\alpha_p, \alpha_d)$
	\STATE $p \gets p + \alpha  \Delta p$
 	\STATE $y \gets y + \alpha  \Delta y$
  	\STATE $\lambda \gets \lambda + \alpha \Delta \lambda$
\ENDFOR
\STATE $p^* \gets p$
\STATE $\lambda^* \gets \lambda$
\STATE $y^* \gets y$
\end{algorithmic}
\end{algorithm}

\clearpage
\subsection{Merit Functions and Line Search Methods}

After the local Quadratic subproblem, which yields the solution $p^*$ and Lagrange multiplers $\lambda^*$, has been solved we must then use the solution to the local surface to update the best sequence of inputs $\bU^*$. In what follows, a line search method will be used to update the solution $\bU^*$.

Because the local region of the surface is approximated as a quadratic surface, it cannot be guarenteed if this approximation will remain 'good' for the entire step coresponding to the optimal quadratic solution $p^*$.
Merit functions are therefore used to check if the approximation is good enough to allow taking a fraction of the local solution, where this fraction is the value $\alpha\in(0,1)$, and check that enough progress in decreasing the cost function is made by taking the step.
These functions are particularly important in the presence of non-linear dynamics, such as the cart-pendulum problem studied here.

The implementation of a merit function is different for trust region methods, but for line searching methods, failing the merit function test usually results by shortening $\alpha$ by some fraction $\tau$ before reattempting the test, or in the case of tiny $\alpha$ values, aborting the step and declaring a converged solution.

Typically, employing a merit function is functionally identical to the Armijo rule or First Wolfe condition.
When strictly using this rule, is not enough to simply have made progress on the surface to pass the test, but we must have made sufficient progress.
Test can be written mathematically to be
\begin{align}
\label{eq:wolfe1cond}
\phi(\bU^*+\alpha p^*) \leq \phi(\bU^*) + \eta\alpha (p^*)^T\frac{\partial \phi (\bU^*)}{\partial \bU},
\end{align}

where $\eta=(0,1)$ is a tuning parameter with a typical value of $10^{-4}$ \cite{nocedal1999numerical}, $\phi(\cdot)$ is a merit function, and $\bU^*$ is the current best solution about where the QP problem was formed to approximate the local region. 
When the $\eta$ value is this small, the component seems to have little effect on the solution, and may not be worth the computational cost of implementation. Note that an $\eta$ value of zero just requires a step reduces evaluation of the merit function, and has no expectation on by how much.

For the unconstrained problem, setting the merit function to an evalution of the cost surface at the proposed point would be a good choice,
$$\phi(\bU) = V(\mathbf{u}),$$
in which case, the gradient of the local quadratic approximation turns out to be a key component of \eqref{eq:wolfe1cond}, as
$$\frac{\partial \phi (\bU)}{\partial \bU} \Big\rvert_{\bU=\bU^*}= \vec{g}(\bU^*).$$
But NMPC often needs to deal with constraints. 
From the paper, it is suggested to handle constrained NMPC optimisation using a merit function that originates from \cite{nocedal1999numerical},
$$\phi(\bU) = V(\bU)+\frac{1}{\mu} \lVert c(\bU)-y \rVert_1, \text{ where } \mu = \frac{1}{\lVert \lambda^*\rVert_\infty+0.1},$$
where $c(\cdot)$ is an evaluation of the constraints vector, where to satisfy the constraints every component element should be less than or equal to zero $c_i(\bU) \leq 0 \, \forall i$. Note that there is a subtraction of the slack variables which was added to this equation as per equation 17.47 of \cite{nocedal1999numerical} to handle inequality constraints. When solving the QP problem, $c(\cdot)$ is linearised about the local region to calculate $A$ and $b$, where $Ap \geq b$. Note that as an implementation detail, $b = c(\bU^*)$.

The problem with this approach is that we had to give $\mu$ a value such that the constraint violation would dominate the score when constraints are violated and have little affect when the constraints are satisfied.
Additionally, the gradient of inequality constraint improvement is discontinious, as when moving along the step, constraints can be satisfied and no more improvement is to be expected, so the strict scaling of $\alpha$ in \eqref{eq:wolfe1cond} shouldn't be used forinequality merit functions.

Instead of implementing the line search and merit function as per the origional paper, a reasonable argument could be made to separate the merit function indicating a reduction of surface cost from the merit function for a reduction of constraint violation , i.e.,

$$\phi_s(\bU) = V(\bU), \quad \phi_c(\bU) =\lVert c(\bU)\rVert_1^+,$$
where $\lVert v \rVert_1^+$ is a modified norm-1 and returns the sum only positive components in vector $v$.

Typically interior point methods stay in the valid set, but the problem may be overconstrained, for heavily non-linear constraints a feasible solution may take a few iterations to solve, and possible the first iteration of the algorithm with a new $\bU^+$ sequence may not have an input sequence that complies to the constraints.

The expected gradient of the separated merit function $\frac{\partial \phi_c (\bU)}{\partial \bU} \rvert_{\bU=\bU^*}$ is the sum of constraint violated rows in $A$, but these components saturate when moving past the constraint boundary.

\begin{align}
\label{eq:costgrad45} 
\bar{v}(A,b, p^*, \alpha) \triangleq \sum_{i \in \mathcal{I} } \min(b_i, \alpha a_i),\text { where } a = A p^*, \text{ and } \{\mathcal{I} | b_i > 0 \},
\end{align}
where $\mathcal{I}$ is the set of indicies of violated constraints, $a_i$ is the i-th element of $a$, and $b_i$ is the i-th element of $b$.

\begin{algorithm}[!htbp]
\begin{algorithmic}
\caption{Backtracking Line Search}
\label{alg:linesearch}
\REQUIRE Number of backtracking iterations $n_L$, and a choice of tuning parameters  $\eta=(0,1),$ $\eta_c=(0,1)$, and $\bar\tau = (0,1)$.
\REQUIRE Last solution $\bU^*$
\REQUIRE The QP parameters $H(\bU^*)$, $\vec{g}(\bU^*)$, $A(\bU^*)$, $b(\bU^*)$ 
\REQUIRE $\lambda$, and $y$ vectors used to initialise QP solve.
\REQUIRE QP solve proposed step $p^*$, lagrange multipliers $\lambda^*$, and slack variables $y^*$.
\REQUIRE Surface cost evaluation $V(\bU^*)$, which is best obtained when forming the QP problem.
\STATE Set $v_c(\bU^*) \gets b(\bU^*)$ .
\STATE Clear converged flag $\varphi\gets 0$.
\STATE Set $\alpha\gets 1.0$ .
\FOR{$i=1,..., n_L$}
	\STATE $\bar\bU \gets \bU^*+\alpha p^*$.
	\STATE Predict forward with $\bar\bU$ and compute state evolution $x_{2:N+1}(\bar\bU)$.
	\STATE Evaluate constraint function $c(\bar\bU)$,  using $x_{2:N+1}(\bar\bU)$ for any state or state dependant constraints.
	\STATE Compute $\bar{v}(A,b, p^*, \alpha)$ using \eqref{eq:costgrad45}.
	\IF {$v_c(\bar\bU) \leq v_c(\bU^*) -\eta_c \bar{v}(A,b, p^*, \alpha)$}
		\STATE Update the solution $\bU^* \gets \bar\bU$.
		\STATE Update Lagrange multipliers $\lambda \gets \lambda + \alpha ( \lambda^* -  \lambda)$.
		\STATE Update Lagrange multipliers $y \gets y + \alpha ( y^* -  y)$.
		\RETURN
	\ENDIF
	\STATE Using $x_{2:N+1}(\bar\bU)$, evaluate \eqref{eq:evectornn} to compute $e(\bar\bU)$ to allow \eqref{eq:Vsquaredrelationship} to yield $V(\bar\bU)$.
	\IF {$v_c =0$}
		\IF {$V(\bar\bU) \leq V(\bU^*) + \eta\alpha (p^*)^T\bar{g}(\bU^*)$}
			\STATE Update the solution $\bU^* \gets \bar\bU$.
			\STATE Update Lagrange multipliers $\lambda \gets \lambda + \alpha ( \lambda^* -  \lambda)$.
			\STATE Update Lagrange multipliers $y \gets y + \alpha ( y^* -  y)$.
			\RETURN
		\ENDIF
	\ENDIF
	\STATE Shrink the step size $\alpha \gets \bar\tau \alpha$.
\ENDFOR
\STATE Set converged flag $\varphi\gets 1$.
\end{algorithmic}
\end{algorithm}

\clearpage
\section{Problem Setup}

Quadratic cost chosen in equation 3 of the paper
$$V(\mathbf{u})=\sum_{k=1}^N x_{k+1}^T Q x_{k+1} + u_k^T R u_k,$$
where the prediction of state $x$ is a function of $u_{1:N}$.

This is factorised as 
\begin{align} 
V(\mathbf{u}) = e(\mathbf{u})^Te(\bU) \label{eq:Vsquaredrelationship} 
\end{align}
 in equation eleven, where
\begin{align}
\label{eq:evectornn}
e(\mathbf{u}) = \begin{bmatrix} Q^{1/2}x_2 \\ \dots \\ Q^{1/2}x_{N+1}\\R^{1/2}u_1\\ \dots \\ R^{1/2}u_N\end{bmatrix}
\end{align}

What follows is very interesting. In the papers equation 12

\begin{align}
e(\mathbf{u}+p) \approx e(\mathbf{u}) + J(\mathbf{u})p,
\end{align}

by approximating $e(\mathbf{u})$ with a linear surface, in accordance with \eqref{eq:Vsquaredrelationship}, we are effectively approximating the local region of $V(\mathbf{u})$ with a quadratic surface.

What the paper doesn't exadurate, is the nice closed form equations for the gradient and hessian of the surface which can be obtained from this approximation.

Begin with
$$V(\bU +p) =e(\bU +p)^Te(\bU +p)=p^TJ(\bU)^TJ(\bU)p+p^TJ(\bU)^Te(\bU)+e(\bU)^TJ(\bU)p + e(\bU)^Te(\bU)$$

as scalar terms can be transposed without effect $p^TJ(\bU)^Te(\bU)=e(\bU)^TJ(\bU)p$, and therefore

$$V(\bU +p) =e(\bU +p)^Te(\bU +p)=p^TJ(\bU)^TJ(\bU)p+2e(\bU)^TJ(\bU)p + e(\bU)^Te(\bU)$$

By equating $V(\cdot)=2y(\cdot)$ from \eqref{eq:quadraticbase}, we obtain the parameterisation of the surface with
\begin{align}
\label{eq:gradandhessian}
H(\bU)=J(\bU)^TJ(\bU), \text{ and } \vec{g}(\bU)=J(\bU)^Te(\bU),
\end{align}
where $H(\bU)$ is given in the paper in equation 15 but $\vec{g}(\bU)$ is absent.

It could be concluded that at this stage, if $e(\bU)$ and $J(\bU)$ are known, we can form such a quadratic surface to optimise over.

\subsection{Forming the Jacobian}
The remaining task of computing $J(\mathbf{u})$ is largely left for the reader, but in accordance with the first order Taylor series expansion,
\begin{align}
J(\mathbf{u}) = \frac{\partial e(\bU) }{\partial \bU^T}= \begin{bmatrix} Q^{1/2} \frac{\partial}{\partial \bU^T}x_2 \\ \vdots \\ Q^{1/2} \frac{\partial}{\partial \bU^T}x_{N+1}\\R^{1/2} \frac{\partial}{\partial \bU^T}u_1\\ \vdots \\ R^{1/2} \frac{\partial}{\partial \bU^T}u_N\end{bmatrix}
\end{align}

The lower part of $J$ is trivial to derive as 
\begin{align}
\label{eq:inputisdelta}
\frac{\partial u_k}{\partial u_j} = \begin{cases} 
      1 & k=j \\
     0 & k \neq j
   \end{cases}
\end{align}
this effectively turns the lower part of $J$ into a block diagonal matrix, and if $R$ is chosen to be a diagonal matrix, so too will the lower part of $J$.

The upper part of $J$ is a much more involved computation.
From inspection of the discretised process model $x_{k+1} = f(x_k,u_k)$ it is apparent that such model is causal, and in input in the future that has not happened yet has no bearning on the current state, or to put more formally

\begin{align}
\frac{\partial x_k}{\partial u_j} = 0 \quad \text{if } k \leq j,
\end{align}
which yields the Jacobian in the paper

\begin{align}
J(\bU)=\begin{bmatrix}
Q^{1/2}\frac{\partial x_2}{\partial u_1^T} & 0 & 0 & \dots & 0\\
Q^{1/2}\frac{\partial x_3}{\partial u_1^T} & Q^{1/2}\frac{\partial x_3}{\partial u_2^T} & 0 & \dots & 0\\
\vdots &&&&\vdots\\
Q^{1/2}\frac{\partial x_{N+1}}{\partial u_1^T} & Q^{1/2}\frac{\partial x_{N+1}}{\partial u_2^T} & \dots & \dots & Q^{1/2}\frac{\partial x_{N+1}}{\partial u_N^T}\\
R^{1/2}&0&0&\dots&0\\
\vdots &&&&\vdots \\
0&\dots&\dots&0&R^{1/2}
 \end{bmatrix}.
\label{eq:Jacobianmain}
\end{align}

Here we take a moment to appreciate that since $H(\bU) = J(\bU)^TJ(\bU)$, the $R$ penalty plays an important role in maintaining positive-definiteness of the Hessian $H(\bU)$, and by selecting $R$ carefully\footnote{For example a suitable $R$ could be a diagonal matrix with positive elements along the diagonal, which are large enough to avoid numerical problems, e.g. $\geq 10^{-6}$.}, the hessian is not semi-positive definite. This ensures that the Hessian is invertable, which is important for computing the search direction as we will see later. Okay, moment's over.

When constructing $J$, the original paper provides a helpful hint that many of the terms composing the top of $J$ are a function of components in the row above,

\begin{align}
\frac{\partial x_{k+1}}{\partial u_j^T} = \frac{\partial f(x_k, u_k)}{\partial x_k^T}\frac{\partial x_k}{\partial u_j^T} + \frac{\partial f(x_k, u_k)}{\partial u_k^T}\frac{\partial u_k}{\partial u_j^T} \label{eq:chainrulemasterwe}
\end{align}

at this step, it is useful to remember the result of \eqref{eq:inputisdelta}, and that $\frac{\partial x_k}{\partial u_j^T} $ is the term above the term currently being computed in matrix $J$ without the $Q^{1/2}$ component.

The paper continues to discuss the continious time process model $g(x_k,u_k)$ and the integration method used to convert from this model to discrete time as $f(x_k,u_k)$ but, most likely due to brevity, does not provide the reader with equations to do this.
Such a step and the algorithms needed is common to implementing MPC on any real-time dynamic system, and as such the detail is provided in following sections~\ref{sec:discretisationmethod} and \ref{sec:discretisationapproach}.
But for now, we continue to setup the QP problem with a short section on inequality constraints.

\subsection{Inequality Matrices}

SQP solvers typically require a linearisation of the constraints function $0 \geq c(\mathbf{u})$ to the form the problem
$$Ap \geq b,$$
where $b(\bU^*)=c(\mathbf{u}^*)$ and for MPC, the input being considered by the QP is $\mathbf{u} = \mathbf{u}^* + p$ and constraints on limited input effort such as $b_l \leq u_k \leq b_u$ for $k=1,2, \dots$ are easily captured in this form.
Lower input bounds can be written as
$$b_l \leq \mathbf{u} = \mathbf{u}^* + p, \quad b_l -\bU^* \leq p,$$
and similarly upper input bounds written as
$$b_u \geq \mathbf{u} = \mathbf{u}^* + p, \quad b_u -\bU^* \geq p, \quad -b_u +\bU^* \leq -p,$$
or consicely written for the entire input set as
\begin{align}
\label{eq:inputconstraint}
\begin{bmatrix}
\mathbf{I}_{\bar{n},\bar{n}}\\
-\mathbf{I}_{\bar{n},\bar{n}}
\end{bmatrix}
p \geq
\begin{bmatrix}
-u^*_1 + b_l \\
\vdots \\
-u^*_N + b_l\\
u^*_1-b_u\\
\vdots \\
u^*_N-b_u 
\end{bmatrix},
\end{align}

where $\bar{n}=Nn_u$ is the number of input variables that are being solved for.

Constraints on the state such as a cart translation limits are unfortunately not linear, and require more effort.
Thankfully such a constraint can make use of many previously computed quantities
First we take a linearisation of the state about the current optimal sequence $\mathbf{u}^*$,

$$x_k(\mathbf{u}) = \frac{\partial x_k}{\partial \mathbf{u}^T}(\mathbf{u}-\mathbf{u}^*) +x(\mathbf{u}^*) 
= \frac{\partial x_k}{\partial \mathbf{u}^T}p +x(\mathbf{u}^*),$$

so for an lower bound
$$ b_l^x \leq x_k(\mathbf{u})=   \frac{\partial x_k}{\partial \mathbf{u}^T}p +x(\mathbf{u}^*) ,$$
$$\frac{\partial x_k}{\partial \mathbf{u}^T}p \geq  b_l^x   - x(\mathbf{u}^*) .  $$

Now the entire input sequence can be written as

\begin{align}
\label{eq:statelowerbound}
\begin{bmatrix}
\frac{\partial x_2}{\partial u_1^T} & 0 & 0 & \dots & 0\\
\frac{\partial x_3}{\partial u_1^T} & \frac{\partial x_3}{\partial u_2^T} & 0 & \dots & 0\\
\vdots &&&&\vdots\\
\frac{\partial x_{N+1}}{\partial u_1^T} & \frac{\partial x_{N+1}}{\partial u_2^T} & \dots & \dots &\frac{\partial x_{N+1}}{\partial u_N^T}
 \end{bmatrix}
p \geq
\begin{bmatrix}
\vec{b}^x_l - x_2\\
\vec{b}^x_l - x_3\\
\vdots \\
\vec{b}^x_l -x_{N+1}
\end{bmatrix}
,
\end{align}

Upper bounds can also be written as

$$ b_u^x \geq x_k(\mathbf{u})=   \frac{\partial x_k}{\partial \mathbf{u}^T}p +x(\mathbf{u}^*),$$
$$- b_u^x \leq - \frac{\partial x_k}{\partial \mathbf{u}^T}p - x(\mathbf{u}^*),$$
$$-\frac{\partial x_k}{\partial \mathbf{u}^T}p \geq  -b_u^x  + x(\mathbf{u}^*).   $$

Now the entire input sequence can be written as
\begin{align}
\label{eq:stateupperbound}
-\begin{bmatrix}
\frac{\partial x_2}{\partial u_1^T} & 0 & 0 & \dots & 0\\
\frac{\partial x_3}{\partial u_1^T} & \frac{\partial x_3}{\partial u_2^T} & 0 & \dots & 0\\
\vdots &&&&\vdots\\
\frac{\partial x_{N+1}}{\partial u_1^T} & \frac{\partial x_{N+1}}{\partial u_2^T} & \dots & \dots &\frac{\partial x_{N+1}}{\partial u_N^T}
 \end{bmatrix}
p \geq
-
\begin{bmatrix}
\vec{b}^x_u - x_2\\
\vec{b}^x_u- x_3\\
\vdots \\
\vec{b}^x_u -x_{N+1}
\end{bmatrix}
.
\end{align}

Hopefully it is recognised that \eqref{eq:statelowerbound} and \eqref{eq:stateupperbound} share many terms with the Jacobian \eqref{eq:Jacobianmain}, which is already required to be generated.
Finally, note that not all states require constraints, and only the needed rows should be generated. The required subset of constraints from \eqref{eq:inputconstraint}, \eqref{eq:statelowerbound} and \eqref{eq:stateupperbound} can be combined to form a single set of constraint matrices of $A(\mathbf{u}^*)$ and $b(\mathbf{u}^*)$ such that $A(\mathbf{u}^*)p \geq b(\mathbf{u}^*)$.

\subsection{Discretisation Method}
\label{sec:discretisationmethod}
The paper considers integration of the continious system by Euler's method, where each discrete timestep is $\Delta$ seconds apart, and is further broken into $M$ equally spaced parts in time to conduct the integration
$$\delta = \frac{\Delta}{M}$$

Note that while increasing M will offer more accurate integration, this comes at drastically increased computational cost of the overall algorithm, and this may result in the control action being computed after it was due.

We employ the following notation for these intermediate steps, $x_{i/k}$ is the result from the i-th completed intermediate step completed on $x_k$, and if there are indeed $M$ intermediate steps, then
$$x_{M/k} = x_{0/k+1} = x_{k+1}.$$

The Euler integration for an arbibary intermediate step is therefore
\begin{align}x_{i+1/k} = x_{i/k} + \delta g( x_{i/k}, u_k), \label{eq:eulersubstep}
\end{align}
where $\dot{x} = g(x,u)$ is the continious process model function.

\subsection{Discretisation Approach}
\label{sec:discretisationapproach}
Note that the following process could be made time-variant with little effort, and the corresponding code is written to support this.
For brevity, time variance of the process model is excluded in this section.
Additionally, in order to make the following more readable, we use the shorthand of
\begin{align}
g_u(x, u) = \frac{\partial g(x, u)}{\partial u^T}, \, \text{and } g_x(x, u) = \frac{\partial g(x, u)}{\partial x^T},
\end{align}

as the partial derivatives of the continious time process model wrt current state and input are actively used.

To construct \eqref{eq:Jacobianmain} using \eqref{eq:chainrulemasterwe}, we require two terms $\frac{\partial f(x_k, u_k)}{\partial x_k^T}$ and $\frac{\partial f(x_k, u_k)}{\partial u_k^T}$.  There are a few ways these could be derived.
One way way might be to perform the integration steps using a symbolic toolbox, and taking the derivative of the function with respect to the variable of interest.
Another, more general, way would be derive the equation to propagate the contributions forwards in time, as follows.

To derive these terms we being by taking the derivates of \eqref{eq:eulersubstep} wrt $ x_{i/k}$ and $u_k$. These yield
\begin{align} \frac{\partial x_{i+1/k}}{\partial x^T_{i/k} } = \mathbf{I} + \delta g_x( x_{i/k}, u_{i/k}) ,\\
\frac{\partial x_{i+1/k}}{\partial u^T_{i/k} } = \delta g_u( x_{i/k}, u_{i/k}) .
\end{align}

For one discrete time step

\begin{align}
\frac{\partial x_{k+1}}{\partial x^T_{k}}=\frac{\partial x_{M/k}}{\partial x^T_{0/k}} &= 
\frac{\partial x_{M/k}}{\partial x^T_{M-1/k}} 
\dots
\frac{\partial x_{2/k}}{\partial x^T_{1/k}}
\frac{\partial x_{1/k}}{\partial x^T_{0/k}} \nonumber \\
&=
(\mathbf{I} + \delta g_x( x_{M-1/k}, u_k))
(\mathbf{I} + \delta g_x( x_{M-2/k}, u_k))
\dots 
(\mathbf{I} + \delta g_x( x_{0/k}, u_k)),
\end{align}

\begin{align}
\frac{\partial x_{k+1}}{\partial u^T_{k}}=\frac{\partial x_{M/k}}{\partial u^T_{k}} = &\frac{\partial x_{M/k}}{\partial u^T_{0/k}} + \dots+\frac{\partial x_{M/k}}{\partial u^T_{M-2/k}}+\frac{\partial x_{M/k}}{\partial u^T_{M-1/k}}   \nonumber \\
 =& 
\frac{\partial x_{M/k}}{\partial x^T_{M-1/k}} 
\dots
\frac{\partial x_{2/k}}{\partial x^T_{1/k}}
\frac{\partial x_{1/k}}{\partial u^T_{0/k}} \nonumber  \\
&+ \frac{\partial x_{M/k}}{\partial x^T_{M-1/k}} 
\dots
\frac{\partial x_{3/k}}{\partial x^T_{2/k}}
\frac{\partial x_{2/k}}{\partial u^T_{1/k}} \nonumber  \\
&+ \dots  \nonumber \\
&+ \frac{\partial x_{M/k}}{\partial u^T_{M-1/k}} \nonumber  \\
=& 
(\mathbf{I} + \delta g_x( x_{M-1/k}, u_k)) \dots (\mathbf{I} + \delta g_x( x_{1/k}, u_k)) (\delta g_u( x_{0/k}, u_k) ) \nonumber \\
&+ (\mathbf{I} + \delta g_x( x_{M-1/k}, u_k)) \dots (\mathbf{I} + \delta g_x( x_{2/k}, u_k)) (\delta g_u( x_{1/k}, u_k) ) \nonumber  \\
&+ (\mathbf{I} + \delta g_x( x_{M-1/k}, u_k)) \dots (\mathbf{I} + \delta g_x( x_{3/k}, u_k)) (\delta g_u( x_{2/k}, u_k) )   \nonumber \\
&+ \dots \nonumber  \\
&+ (\delta g_u( x_{M-1/k}, u_k) ).
\end{align}

We can therefore compose a straight-forward algorithm to conduct this discretisation and provide us with the gradients needed to construct the Jacobian, and form the quadratic surface to be optimised over, outlined in Algorithm~\ref{alg:discretisecontssystem}.
\begin{algorithm}[!htbp]
\begin{algorithmic}
\caption{Euler Segment Integration with Gradients}\label{alg:discretisecontssystem}
\REQUIRE Availability of the functions $g_u(x, u)$, $g_x(x, u)$, and $g(x, u)$ for evaluation.  {In practice, a single function could evaluate all three.}
\REQUIRE Integer $M \geq 1$ . {This is the number of integration slices that are taken between discrete timesteps.}
\STATE $\delta \gets \Delta / M$
\STATE $x \gets x_{k}$
\STATE $u \gets u_k$ . {Input signal is assumed to be held constant over the discrete time step.}
\STATE $X \gets \mathbf{I}_{nx, nx}$
\STATE $U \gets \mathbf{0}_{nx, nu}$
\FOR{$i= 0,...,M-1$}
	\STATE $\bA \gets  \mathbf{I}_{nx,nx} + \delta g_x(x,u)$
	\STATE $X \gets \bA X$
	\STATE $U \gets \bA U+ \delta  g_u(x, u)$
	\STATE $x \gets x + \delta g(x, u)$
\ENDFOR

\STATE $\frac{\partial f(x_k, u_k)}{\partial x_k^T} \gets X$. These value could also be useful for implementing an EKF filter. 
\STATE $\frac{\partial f(x_k, u_k)}{\partial u_k^T} \gets U$
\STATE $x_{k+1} \gets x$
\end{algorithmic}
\end{algorithm}

\clearpage

\section{Algorithm Overview}
With all working components now introduced, the basic MPC algorithm can now be summarised in Algorithm~\ref{alg:MPCoverview}. This algorithm could be considered to be Naive, as it computes large matricies, including a large symetric Hessian matrix $H$.
Details on an improved implementation will be provided in the following section, but it is important to stop here and grasp the many components and their interactions.

\begin{algorithm}[!htbp]
\begin{algorithmic}
\caption{Naive NMPC Algorithm Iteration}\label{alg:MPCoverview}
%
\REQUIRE The continious plant model $g(x, u)$ with the derivatives $g_u(x, u)$, and $g_x(x, u)$ for evaluation.
\STATE Set $\mathbf{u}^* \gets \mathbf{u}^+$ with the hot start input sequence.
\STATE Set  $y \gets  \vec{\mathbf{1}}_m$, and $\lambda \gets  \vec{\mathbf{1}}_m$, where $m$ is the number of inequality constraints.
\FOR{$i= 0,...$}
	\FOR{$k=1, ..., N$}
		\STATE Using $\mathbf{u}^*$ and Algorithm~\ref{alg:discretisecontssystem}, evaluate the state trajectory $x_{k+1}$ and derivatives $\frac{\partial f(x_k, u_k)}{\partial x_k^T}$and $\frac{\partial f(x_k, u_k)}{\partial u_k^T}$. 
		\STATE Construct part of $J(\mathbf{u}^*)$ using \eqref{eq:Jacobianmain}, $e(\mathbf{u}^*)$ using \eqref{eq:evectornn}.
		\STATE Construct part of $A(\mathbf{u}^*)$ and $b(\mathbf{u}^*)$ using  \eqref{eq:inputconstraint}, \eqref{eq:statelowerbound} and \eqref{eq:stateupperbound}.
	\ENDFOR
	\STATE Evaluate $H(\bU^*)$, and $\vec{g}(\bU^*)$ using \eqref{eq:gradandhessian} and the values for $J(\mathbf{u}^*)$ and $e(\mathbf{u}^*)$.
	\STATE Using Algorithm~\ref{alg:MetPredCorr}, solve the Quadratic Program for $p^*$, $\lambda^*$, and $y^*$.
	\STATE Using Algorithm~\ref{alg:linesearch}, update $\mathbf{u}^*$, $\lambda$, and $y$, using $p^*$, $\lambda^*$, $y^*$ and retrieve converged flag state $\varphi$.
	\IF {$\varphi$ =1 }
		\STATE \textbf{break}
	\ENDIF
\ENDFOR
\STATE Allocate $u_1^*$ to the actuators with correct timing.
\STATE Shift the input sequence to compute the hotstart for the next iteration $\mathbf{u}^+ \gets [u^*_{2:N}, u^*_N]$.
\end{algorithmic}
\end{algorithm}

\section{Results}
\begin{figure}[h]
    \centering
    \includegraphics[width=0.7\textwidth]{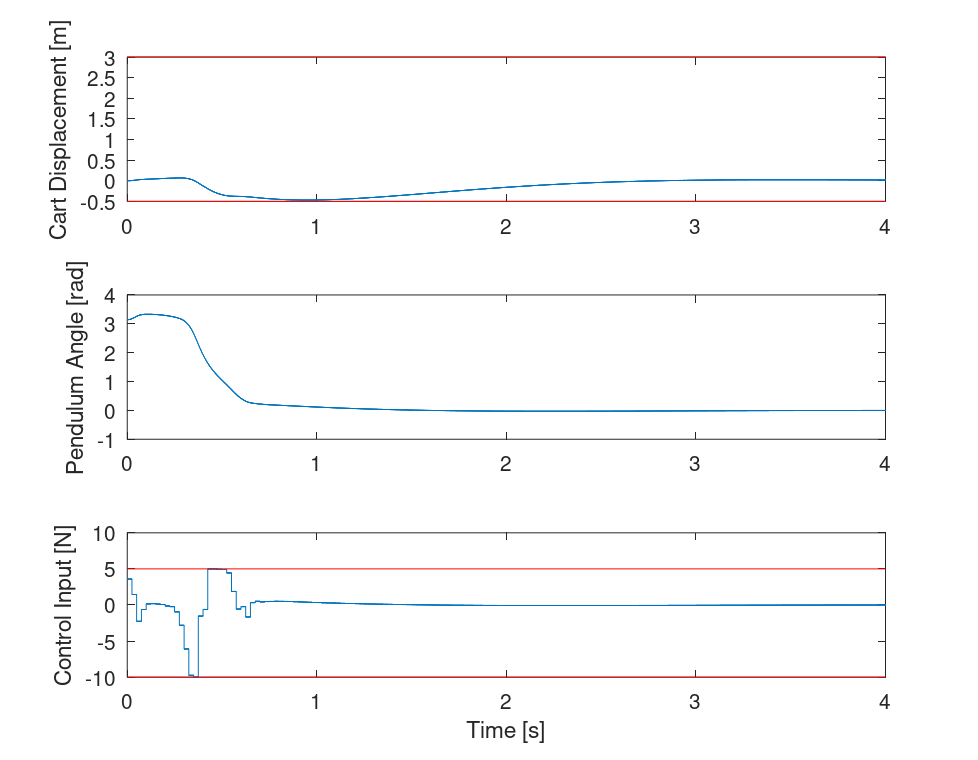}
    \caption{Running the NMPC controller of a cart-pendulum in simulation.}
    \label{fig:simresults}
\end{figure}

As shown by Figure~\ref{fig:simresults}, the algorithm did control the cart-pendulum and regulate to the setpoint of $\theta=p=0$, and it did this while respecting the in-equality constraints.
The required run-time performance was not achieved however, as the average computational time was 0.6s\footnote{PC used was running Intel(R) Core(TM) i7-7820HK CPU @ 2.90GHz with 32GB of RAM.}, falling short of the target of under 25ms. This is most likely due to octave interpreter being much slower than compiled C code. 

Further work on this paper may involve an embedded implementation focusing on real-time performance.

\clearpage
\appendix
 \section{Analytical Mechanics}
I ran out of letters!
$g$ in this section is the constant for acceleration due to gravity, which nominally has the value of $g=9.8$.
This derivation uses a different friction model to the origional paper.
$b$ and $c$ are also overloaded, and in this section they refer to viscous damping co-efficients of the cart and pendulum respectively.
$\tau$ in this section is a torque, not a tuning constant for the QP solver.
$\mathcal{L}$ in this section is a different type of Lagrangian to that used for Lagrangian multipliers.

We also make use of the shorthand
$\dot{\theta} = \omega$, $\ddot{\theta} = \dot{\omega} = \alpha,$
 and 
$\dot{p} = v$, $\ddot{p}=\dot{v}=a$.

\begin{figure}[h]
    \centering
    \includegraphics[width=0.7\textwidth]{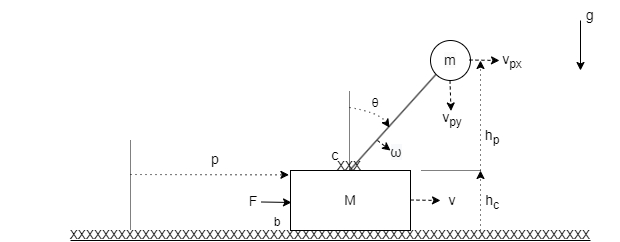}
    \caption{The cart pendulum problem.}
    \label{fig:cartpenddiag}
\end{figure}

We begin with the modelling diagram is shown in Figure~\ref{fig:cartpenddiag}.

Then $h_p = \ell \cos(\theta)$, $v_{px}=\ell \omega \cos (\theta)$, and $v_{py}=\ell \omega \sin (\theta)$.

\subsection{$v_p$ components}
The velocity of the pendulum mass relative to the inertial reference frame, $v_p$ can be expressed as
\begin{align}
v_p^2 &= (v+\ell \omega\cos (\theta) )^2 +(\ell \omega \sin (\theta))^2 \nonumber \\
&= v^2 + 2\ell\omega v \cos (\theta) +\ell^2\omega^2\cos(\theta)^2 +\ell^2\omega^2\sin(\theta)^2 \nonumber \\
&= v^2 + 2\ell\omega v \cos(\theta) + \ell^2 \omega^2(\cos(\theta)^2 + \sin(\theta)^2) \nonumber \\
&=v^2  + 2\ell \omega v \cos (\theta) + \ell^2\omega^2,
\end{align}
as $\cos(\theta)^2 + \sin(\theta)^2 = 1$.

\subsection{Lagrangian Formulation}

\begin{align}
\mathcal{T} = \frac{1}{2}Mv^2+ \frac{1}{2} m v_p^2
\end{align}

\begin{align}
\mathcal{V} = mgh_p + h_c= mg\ell \cos (\theta) + h_c
\end{align}

\begin{align}
\mathcal{L} &= \mathcal{T} - \mathcal{V} \nonumber \\
&= \frac{1}{2}Mv^2 + \frac{1}{2}m(v^2 + 2\ell\omega v\cos (\theta) + \ell^2\omega^2) - mg\ell \cos (\theta) - h_c. \label{eq:lagrangianmain}
\end{align}

Now generate equations for each generalised coordinate.

\clearpage
\subsubsection{$\theta$ Generalised Coordinate Equation}
\begin{align}
\frac{d}{dt} \left( \frac{\partial \mathcal{L}}{\partial \omega} \right) - \frac{\partial \mathcal{L}}{\partial \theta} = \tau, \label{eq:thetalagrangegeneralcoordinateeq}
\end{align}
where $\tau$ is equal to applied torques, and in this case is equal to the damping torque $\tau = -c\omega$. The components are derived as follows, from \eqref{eq:lagrangianmain}.

\begin{align}
\frac{\partial \mathcal{L}}{\partial \omega} = m\ell^2 \omega +m\ell v \cos (\theta)
\end{align}
Then using $(uv)' = u'v+uv'$,
\begin{align}
\frac{d}{dt}\frac{\partial\mathcal{L}}{\partial \omega}&=m\ell^2\alpha +m\ell a \cos (\theta) + m \ell v \frac{\partial}{\partial \theta} (\cos (\theta)) \frac{\partial \theta}{\partial t}\nonumber \\
&=m\ell^2\alpha + m\ell a \cos(\omega) - m\omega \ell v \sin(\theta).
\end{align}

We can obtain the final component
\begin{align}
\frac{\partial \mathcal{L}}{\partial \theta} &= -m\ell \omega v \sin (\theta) + mg\ell \sin (\theta)\nonumber \\
&= -m\ell \sin (\theta) (\omega v -g).
\end{align}

Substituting the parts into \eqref{eq:thetalagrangegeneralcoordinateeq} yields
\begin{align*}
m\ell^2 \alpha +m\ell a \cos (\theta) -m\omega\ell v \sin (\theta) +m\ell\sin(\theta)(\omega v-g)=-c\omega
\end{align*}
and after cancelling compoents, finally
\begin{align}
m\ell^2 \alpha +m\ell a \cos (\theta)  -m\ell g \sin(\theta)=-c\omega. \label{eq:rotationeqn11}
\end{align}

\subsubsection{$p$ Generalised Coordinate Equation}
\begin{align}
\frac{d}{dt}\frac{\partial \mathcal{L}}{\partial v} - \frac{\partial \mathcal{L}}{\partial p} = F, \label{eq:pgeneralisedcoordinatelagrangian}
\end{align}
where $F_\text{in} = F-bv$.
Next taking the partial of \eqref{eq:lagrangianmain},
\begin{align}
\frac{\partial \mathcal{L}}{\partial v} = Mv+mv + m\ell\omega \cos (\theta)
\end{align}

then using $(uv)' = u'v+uv'$
\begin{align}
\frac{d}{dt}\frac{\partial \mathcal{L}}{\partial v} &= Ma+ma + m\ell \alpha \cos (\theta) +m\ell \omega \frac{\partial}{\partial \theta} (\cos (\theta))\frac{\partial \theta}{\partial t} \nonumber \\
&= (M+m)a + m\ell  \alpha \cos(\theta) -m\ell\omega^2 \sin (\theta).
\end{align}

This one was easy
\begin{align}
\frac{\partial \mathcal{L}}{\partial p} = 0.
\end{align}

Finally, by substituting the components into \eqref{eq:pgeneralisedcoordinatelagrangian},
\begin{align}
(m+M)a+m\ell\alpha \cos (\theta) - m\ell \omega^2 \sin (\theta)=F-bv \label{eq:translationeqn11}
\end{align}

\subsubsection{Independant Dynamic Equations}
Begin with the equations \eqref{eq:rotationeqn11} and \eqref{eq:translationeqn11}
\begin{align}
\begin{bmatrix}
M+m & m\ell\cos (\theta) \\
m\ell\cos (\theta) & m\ell^2
\end{bmatrix}
\begin{bmatrix}
\ddot{p}\\ \ddot{\theta}
\end{bmatrix}
=
\begin{bmatrix}
F+m\ell \dot{\theta}^2\sin (\theta) - b\dot{p} \\
m\ell g \sin (\theta) -c\dot{\theta}
\end{bmatrix}
\end{align}

Using
$$\begin{bmatrix} A & B\\C & D\end{bmatrix}^{-1} = \frac{1}{AD-BC}\begin{bmatrix} D & -B\\-C & A\end{bmatrix},$$
then the State-Space equations become

\begin{align}
x=\begin{bmatrix} \dot{p} \\ p \\ \dot{\theta} \\ \theta \end{bmatrix}, \quad \dot{x}=g(x,u)=\begin{bmatrix} \ddot{p} \\ \dot{p} \\ \ddot{\theta} \\ \dot{\theta} \end{bmatrix}=\begin{bmatrix} \ddot{p}(x, u) \\ \dot{p} \\ \ddot{\theta}(x, u) \\ \dot{\theta} \end{bmatrix}, 
\quad u = 
\begin{bmatrix}F\end{bmatrix}
\end{align}

where

\begin{align}
\ddot{p}(x, u) = \frac{\beta_p (x, u)}{\psi(\theta)}= \frac{m\ell^2(F + m\ell\dot{\theta}^2\sin (\theta) - b\dot{p})-m\ell\cos (\theta)  (m\ell g \sin (\theta) - c\dot{\theta})}{(M+m)m\ell^2 -m^2\ell^2 \cos(\theta)^2}
\end{align}

\begin{align}
\ddot{\theta}(x, u) =\frac{\beta_\theta(x, u)}{\psi(\theta)}= \frac{-m\ell \cos (\theta)(F + m\ell\dot{\theta}^2\sin (\theta) - b\dot{p})+ (M+m)  (m\ell g \sin (\theta) - c\dot{\theta})}{(M+m)m\ell^2 -m^2\ell^2 \cos(\theta)^2}
\end{align}

For convience in the following derivations we define

$$ \psi(\theta) \triangleq (M+m)m\ell^2 -m^2\ell^2 \cos^2(\theta)$$

$$\beta_p(x, u) \triangleq m\ell^2(F + m\ell\dot{\theta}^2\sin (\theta) - b\dot{p})-m\ell\cos (\theta)  (m\ell g \sin (\theta) - c\dot{\theta}) $$

$$\beta_\theta(x, u) \triangleq -m\ell \cos (\theta)(F + m\ell\dot{\theta}^2\sin (\theta) - b\dot{p})+ (M+m)  (m\ell g \sin (\theta) - c\dot{\theta})$$

\subsection{Derivatives of cart acceleration, $\ddot{p}$}

$$\frac{\partial \ddot{p}}{\partial F} = \frac{m\ell^2}{(M+m)m\ell^2-m^2\ell^2\cos(\theta)^2}=\frac{1}{M+m(1-\cos(\theta)^2)}$$

$$\frac{\partial \ddot{p}}{\partial \dot{p}}=\frac{-bm\ell^2}{\psi (\theta)}$$

$$\frac{\partial \ddot{p}}{\partial p} = 0$$ 

$$\frac{\partial \ddot{p}}{\partial \dot{\theta}} = \frac{2m^2\ell^3\dot{\theta}\sin (\theta) +cm\ell \cos (\theta) }{\psi(\theta)}$$

Computing $\frac{\partial \ddot{p}}{\partial \theta}$ is eased by deriving the following
$$\frac{\partial \psi(\theta)}{\partial \theta } = -2m^2\ell^2 \cos (\theta) \frac{\partial}{\partial \theta}\cos (\theta)=2m^2\ell^2\cos (\theta)\sin(\theta )$$

\begin{align}\frac{\partial \psi(\theta)^{-1}}{\partial \theta}  &= -\psi(\theta)^{-2}\frac{\partial \psi (\theta)}{\partial \theta}
=\frac{-2 m^2\ell^2\cos(\theta)\sin(\theta)}{\psi(\theta)^{2}}
=\frac{-2 m^2\ell^2\cos(\theta)\sin(\theta)}{(m\ell^2(M+m-m\cos(\theta)^2))^2} \nonumber \\
&= \frac{-2\cos (\theta)\sin (\theta)}{\ell^2(M+m -m \cos (\theta)^2)^2}
=\frac{-\sin (2\theta)}{\ell^2(M+m-m\cos(\theta)^2)^2}
\end{align}

because of the double angle formula $2\cos(\theta)\sin(\theta) = \sin(2\theta)$.

Next by using the chain rule $(uv)' = u'v+uv'$, and $\cos(\theta)^2 + \sin (\theta)^2=1$ and therefore $\sin(\theta)^2=1-\cos(\theta)^2$,

$$\frac{\partial}{\partial \theta} \cos (\theta)\sin (\theta)= -\sin(\theta)^2+\cos(\theta)^2=-1+\cos(\theta)^2+\cos(\theta)^2=2\cos(\theta)^2-1=\cos(2\theta),$$
because of the double angle formula $2\cos(\theta)^2-1=\cos(2\theta)$.

$$\frac{\partial \beta_p (x, u)}{\partial \theta} =m^2\ell^3\dot\theta^2\cos (\theta)-m\ell c \dot\theta \sin (\theta)-m^2\ell^2g\frac{\partial}{\partial \theta}\cos(\theta)\sin(\theta)
=m^2\ell^3\dot\theta^2\cos (\theta)-m\ell c \dot\theta \sin (\theta)-m^2\ell^2g\cos(2\theta)$$

Resuming, with use of the chain rule
$$\frac{\partial \ddot{p}}{\partial \theta}= \frac{\partial}{\partial \theta}\beta_p(x, u)\psi(\theta)^{-1}= \psi(\theta)^{-1}\frac{\partial \beta_p (x, u)}{\partial \theta} + \beta_p (x, u) \frac{\partial \psi(\theta)^{-1}}{\partial \theta}$$

values of the components can be calculated from the above expressions and substituted.

\subsection{Derivatives of pendulum angular acceleration, $\ddot{\theta}$}

$$\frac{\partial \ddot\theta}{\partial F} = \frac{-m\ell \cos (\theta)}{\psi (\theta)}$$

$$\frac{\partial \ddot\theta}{\partial \dot p}=\frac{m\ell b \cos (\theta)}{\psi (\theta)} $$

$$\frac{\partial \ddot \theta}{\partial p} = 0$$

$$\frac{\partial \ddot\theta}{\partial\dot \theta}=\frac{-2m^2\ell^2\dot\theta\cos(\theta)\sin(\theta)-c(M+m)}{\psi(\theta)}=\frac{-m^2\ell^2\dot\theta\sin(2\theta)-c(M+m)}{\psi(\theta)}$$

And finally

$$\frac{\partial \ddot \theta}{\partial \theta} = \frac{\partial}{\partial \theta}\beta_\theta(x, u)\psi(\theta)^{-1}
 = \psi(\theta)^{-1}\frac{\partial \beta_\theta (x, u)}{\partial \theta} + \beta_\theta (x, u) \frac{\partial \psi(\theta)^{-1}}{\partial \theta}$$
where
\begin{align}\frac{\partial \beta_\theta (x, u)}{\partial \theta} &= m\ell \sin (\theta) (F+m\ell\dot\theta^2\sin(\theta)-b\dot p) -m^2\ell^2\dot\theta^2\cos(\theta)^2+(M+m)m\ell g \cos (\theta) \nonumber\\
&= m\ell \sin (\theta) (F+m\ell\dot\theta^2\sin(\theta)-b\dot p) -(m\ell\dot\theta\cos(\theta))^2+(M+m)m\ell g \cos (\theta).\end{align}

\subsection{Derivatives of cart velocity, $\dot{p}$}
For the sake of completeness
$$\frac{\partial \dot p}{\partial F} =0, \frac{\partial \dot p}{\partial \theta} =0, \frac{\partial \dot p}{\partial \dot\theta} =0, \frac{\partial \dot p}{\partial p} =0, \frac{\partial \dot p}{\partial \dot p} =1.$$

\subsection{Derivatives of pendulum angular velocity, $\dot{\theta}$}
For the sake of completeness
$$\frac{\partial \dot \theta}{\partial F} =0, \frac{\partial \dot \theta}{\partial \theta} =0, \frac{\partial \dot \theta}{\partial \dot\theta} =1, \frac{\partial \dot \theta}{\partial p} =0, \frac{\partial \dot \theta}{\partial \dot p} =0.$$

\subsection{Gradient Functions}
With the state and input vectors
$$x=\begin{bmatrix} \dot{p} \\ p \\ \dot{\theta} \\ \theta \end{bmatrix},\quad u = 
\begin{bmatrix}F\end{bmatrix},$$

then the quantities in the paper are

$$g_x(x, u) = \frac{\partial g(x, u)}{\partial x^T} = \frac{\partial \dot x}{\partial x^T} = \begin{bmatrix} 
\frac{\partial \ddot{p}}{\partial \dot{p}} & \frac{\partial \ddot{p}}{\partial {p}} & \frac{\partial \ddot{p}}{\partial \dot{\theta}} & \frac{\partial \ddot{p}}{\partial \theta} \\
\frac{\partial \dot{p}}{\partial \dot{p}} & \frac{\partial \dot{p}}{\partial {p}} & \frac{\partial \dot{p}}{\partial \dot{\theta}} & \frac{\partial \dot{p}}{\partial \theta} \\
\frac{\partial \ddot{\theta}}{\partial \dot{p}} & \frac{\partial \ddot{\theta}}{\partial {p}} & \frac{\partial \ddot{\theta}}{\partial \dot{\theta}} & \frac{\partial \ddot{\theta}}{\partial \theta} \\
\frac{\partial \dot{\theta}}{\partial \dot{p}} & \frac{\partial \dot{\theta}}{\partial {p}} & \frac{\partial \dot{\theta}}{\partial \dot{\theta}} & \frac{\partial \dot{\theta}}{\partial \theta} 
\end{bmatrix}= \begin{bmatrix} 
\frac{\partial \ddot{p}}{\partial \dot{p}} & \frac{\partial \ddot{p}}{\partial {p}} & \frac{\partial \ddot{p}}{\partial \dot{\theta}} & \frac{\partial \ddot{p}}{\partial \theta} \\
1&0&0&0\\
\frac{\partial \ddot{\theta}}{\partial \dot{p}} & \frac{\partial \ddot{\theta}}{\partial {p}} & \frac{\partial \ddot{\theta}}{\partial \dot{\theta}} & \frac{\partial \ddot{\theta}}{\partial \theta} \\
0&0&1&0
\end{bmatrix}
,$$

and

$$g_u(x, u) = \frac{\partial g(x, u)}{\partial u^T} =\frac{\partial \dot x}{\partial u^T} =\begin{bmatrix} 
\frac{\partial \ddot{p}}{\partial F}\\
\frac{\partial \dot{p}}{\partial F}  \\
\frac{\partial \ddot{\theta}}{\partial F} \\
\frac{\partial \dot{\theta}}{\partial F} 
\end{bmatrix}=\begin{bmatrix} 
\frac{\partial \ddot{p}}{\partial F}\\
0  \\
\frac{\partial \ddot{\theta}}{\partial F} \\
0 
\end{bmatrix}.$$

\clearpage
 \section{Implementation Notes}
This section uses the shorthand $J=J(\bU)$, $H=H(\bU)$, and $g=g(\bU)$.

When constraints are absent, the Mehrotra Predictor-Corrector Method should not be used.
Instead the solution to the QP is able to be calculated without iterations as $p^* = -H^{-1}g$.

Note that the parameterisation of $H$ and $g$ should not be used when implementing the solution. As $H^{-1}$ appears in Mehrotra Predictor-Corrector Method and the solution to the unconstrained problem, inverting this large square symmetric matrix is not very efficient.

One possible parameterisation which takes advantage of the problem structure includes using $H^{1/2}$, where


Consider a QR-decomposition of the jacobian, $J(\bU)$, then

$$\mathcal{Q}\mathcal{R} = J,$$
and it follows

$$H^{-1}=(J^TJ)^{-1}=(R^TQ^TQR)^{-1}=(R^TR)^{-1}=R^{-1}R^{-T}$$

So a Q-less QR decomposition of $J$ (which can be constructed by updating with one new row at a time) is a suitable way to compute $H^{1/2}$.

i.e.,  begin with
$$\mathcal{R} \gets \begin{bmatrix} R^{1/2} &0&\dots &\dots&0 \\
0 & R^{1/2}&0&\dots&0 \\
\vdots\end{bmatrix}$$
and repeat the calculation
\begin{align}
\label{eq:Hhalfupdate}
\mathcal{R} \gets qr\left(\begin{bmatrix} \mathcal{R} \\ J_k \end{bmatrix} \right), \forall k=1,\dots, N,
\end{align}
where $J_k$ is the $k$-th row of $J$.
Note that $\mathcal{R}$ is upper-triangular, and therefore can be carefully targeted with Given's rotations to reduce computation.

Similarly, initialise
$$\bar{g} \gets \begin{bmatrix} Ru_1\\Ru_2 \\ \vdots \\ Ru_N\end{bmatrix}$$
followed by the repeated calculation
\begin{align}\label{eq:ebarupdate}
\bar{g} \gets \bar{g} + J_k^T(Q^{1/2}x_{k+1}), \forall k=1,\dots,N.
\end{align}

Note that $J_k$  ends with a shinking zeros matrix with increasing $k$, and this can be leveraged to reduce computation.
The final $\bar{g}=g$.

Other implementation details include computing $V(\bU)$ not by the inner product of $e(\bU)^Te(\bU)$, as this involves producing the vector $e(\bU)$, but by summing the components which compose it cumulatively.

\clearpage
\subsection{Mehrotra Solve}
The back solve
 $\Psi^{-1}(\lambda,y) \gamma(p,\lambda, y, \Delta \lambda^\text{aff} , \Delta y^\text{aff} ,\sigma, \mu) $
is particularly computationally heavy as this naievely inverts a large matrix, where
\begin{align*}
\Psi(\lambda,y) = \begin{bmatrix}H & \mathbf{0}_{\bar{n},m} & -A^T\\A&-\mathbf{I}_{m,m} & \mathbf{0}_{m,m}\\\mathbf{0}_{m, \bar{n}} & \diag (\lambda) & \diag (y)\end{bmatrix}.
\end{align*}

Given that the matrix is sparse, computing an analytical solution to this back solve is likely to yield improved performance.
We wish to solve for $\vec{x}$ where

$$\begin{bmatrix}H & \mathbf{0} & -A^T\\A&-\mathbf{I} & \mathbf{0}\\\mathbf{0} & \diag (\lambda) & \diag (y)\end{bmatrix} \vec{x} = \begin{bmatrix} v_1\\v_2\\v_3\end{bmatrix}
$$

In the folowing derivation, we use $\circled{i}$ to denote the i-th row.
We begin with the operation $\circled{1} \gets H^{-1}\circled{1}$, which yields
$$\begin{bmatrix}\mathbf{I} & \mathbf{0} & -H^{-1}A^T\\A&-\mathbf{I} & \mathbf{0}\\\mathbf{0} & \diag (\lambda) & \diag (y)\end{bmatrix} \vec{x} = \begin{bmatrix} H^{-1}v_1\\v_2\\v_3\end{bmatrix}.
$$

Perform $\circled{2}\gets \circled{2} -A\circled{1}$ to yield
$$\begin{bmatrix}
\mathbf{I} & \mathbf{0} & -H^{-1}A^T\\
\mathbf{0}&-\mathbf{I} & AH^{-1}A^T\\
\mathbf{0} & \diag (\lambda) & \diag (y)
\end{bmatrix} \vec{x} = 
\begin{bmatrix}
H^{-1}v_1\\
v_2-AH^{-1}v_1\\
v_3
\end{bmatrix}.
$$

Then $\circled{2}\gets - \circled{2}$ yields
$$\begin{bmatrix}
\mathbf{I} & \mathbf{0} & -H^{-1}A^T\\
\mathbf{0}&\mathbf{I} & -AH^{-1}A^T\\
\mathbf{0} & \diag (\lambda) & \diag (y)
\end{bmatrix} \vec{x} = 
\begin{bmatrix}
H^{-1}v_1\\
AH^{-1}v_1-v_2\\
v_3
\end{bmatrix}.
$$

Next $\circled{3}\gets\circled{3}-\diag(\lambda)\circled{2}$ yields
$$\begin{bmatrix}
\mathbf{I} & \mathbf{0} & -H^{-1}A^T\\
\mathbf{0}&\mathbf{I} & -AH^{-1}A^T\\
\mathbf{0} & \mathbf{0}  & \diag (y) + \diag(\lambda)AH^{-1}A^T
\end{bmatrix} \vec{x} = 
\begin{bmatrix}
H^{-1}v_1\\
AH^{-1}v_1-v_2\\
v_3 - \diag(\lambda)(AH^{-1}v_1-v_2)
\end{bmatrix}.
$$

Performing $\circled{3} \gets (\diag (y) + \diag(\lambda)AH^{-1}A^T)^{-1}\circled{3}$ yields

$$\begin{bmatrix}
\mathbf{I} & \mathbf{0} & -H^{-1}A^T\\
\mathbf{0}&\mathbf{I} & -AH^{-1}A^T\\
\mathbf{0} & \mathbf{0}  & \mathbf{I}
\end{bmatrix} \vec{x} = 
\begin{bmatrix}
H^{-1}v_1\\
AH^{-1}v_1-v_2\\
\zeta
\end{bmatrix},
$$
where $\zeta \triangleq (\diag (y) + \diag(\lambda)AH^{-1}A^T)^{-1}(v_3 - \diag(\lambda)(AH^{-1}v_1-v_2))$. Note that this itself should be a backsolve.

Finally, $\circled{1} \gets \circled{1} + H^{-1}A^T\circled{3}$, and $\circled{2} \gets \circled{2} +AH^{-1}A^T\circled{3}$ yields
$$\mathbf{I}\vec{x} =\vec{x} = 
\begin{bmatrix}
\vec{x}_1\\ \vec{x}_2\\ \vec{x}_3
\end{bmatrix}
=
\begin{bmatrix}
H^{-1}v_1 + H^{-1}A^T\zeta\\
AH^{-1}v_1-v_2 + AH^{-1}A^T\zeta\\
\zeta
\end{bmatrix}
=
\begin{bmatrix}
H^{-1}v_1 + H^{-1}A^T\zeta\\
A\vec{x}_1-v_2 \\
\zeta
\end{bmatrix}.
$$
\clearpage
\bibliography{references}
\bibliographystyle{chicago}






\end{document}